\documentclass[aps,pra,twocolumn,longbibliography,superscriptaddress,floatfix]{revtex4-1}
\usepackage{amsfonts}
\usepackage{mathtools}
\usepackage{graphicx}
\usepackage{epsfig}
\usepackage{dcolumn}
\usepackage{bm}
\usepackage{amsmath}
\usepackage{graphicx}
\usepackage[latin1]{inputenc}
\usepackage{ulem}
\usepackage{epstopdf}
\usepackage{subfigure}
\usepackage{color}
\usepackage{amsthm}
\usepackage{newlfont}
\usepackage{graphicx}
\usepackage{epstopdf}
\usepackage{appendix}
\usepackage[breaklinks=true]{hyperref}
\usepackage{breakcites}
\usepackage{textcomp}
\usepackage{appendix}
\usepackage{multirow}	
\usepackage{color}
\usepackage{amssymb}
\usepackage{epsfig}
\usepackage{bm}
\usepackage[american]{babel}
\usepackage{braket}
\usepackage{soul}
\usepackage{float}
\hypersetup{colorlinks=true,linkcolor=blue,citecolor=blue,filecolor=blue,urlcolor=blue,pdfstartview=FitH}

\begin{document}
 \title{Thermalization in the one-dimensional Salerno model lattice}
\author{Thudiyangal Mithun}
\affiliation{Department of Mathematics and Statistics, University of Massachusetts, Amherst MA 01003-4515, USA} 
\author{Aleksandra Maluckov}
\affiliation{Vinca Institute of Nuclear Sciences, University of Belgrade, National institute of the Republic of Serbia, P. O. B. 522, 11001 Belgrade, Serbia}
\affiliation{Center for Theoretical Physics of Complex Systems, Institute for Basic Science, Daejeon 34051, Korea} 
\author{Bertin Many  Manda}
\author{Charalampos  Skokos}
\affiliation{Department of Mathematics and Applied Mathematics,University of Cape Town, Rondebosch, 7701 Cape Town, South Africa}
\author{Alan Bishop}
\author{Avadh Saxena}
\affiliation{Center for Nonlinear Studies and Theoretical Division, Los Alamos National Laboratory, Los Alamos, New Mexico 87545, USA}
\author{Avinash Khare}
\affiliation{Department of Physics, Savitribai Phule Pune University, Pune 411007, India}
\author{Panayotis G. Kevrekidis}
\affiliation{Department of Mathematics and Statistics, University of Massachusetts, Amherst MA 01003-4515, USA} 

\begin{abstract}
  The Salerno model constitutes an intriguing interpolation between
  the integrable Ablowitz-Ladik (AL) model and the more standard (non-integrable) discrete
  nonlinear Schr{\"o}dinger (DNLS) one.
The competition of local on-site nonlinearity and nonlinear dispersion
governs the thermalization of this model.
Here, we investigate the statistical mechanics of the Salerno
one-dimensional lattice model in the nonintegrable case and illustrate
the thermalization in the Gibbs regime. 
As the parameter interpolating between the two limits (from DNLS towards AL) 
is varied, the region in the
space of initial energy and norm-densities leading to thermalization expands. 
The thermalization in the
non-Gibbs regime heavily depends on the finite system size; we explore this 
feature
via direct numerical computations for different parametric regimes.
  
\end{abstract}
\maketitle
\pagestyle{myheadings}

\section{Introduction}

Enlightening of the thermalization properties of lattice dynamical
models and complex networks is a crucial issue in understanding and
exploiting the transport and localization phenomena of relevance to a
wide range of physical problems. In spite of substantial research, the coexistence of diverse physical processes and
correlations among them developing on different space-time scales, 
lacks a conclusive interpretation
\cite{Aubry:1975,jackson1978nonlinearity,Bouchaud:1992,Casetti1997FPU,Giardina:2000,Johansson:2004,Bel:2006,Iubini:2013,Biroli:2017,Buonsante:2017,Danieli:2017}. 
It
is therefore natural to expect that the application of statistical and
thermodynamical approaches and related mixing, ergodicity, and energy
equipartition concepts to such problems is a topic of substantial
ongoing interest. Among the numerous prototypical nonlinear physical examples
are:
the statistics of the Fermi-Pasta-Ulam-Tsingou 
chains \cite{Danieli:2017},
chains of Josephson-junctions \cite{Mithun:2019}, 
Gross-Pitaevskii/discrete nonlinear Schr\"odinger
lattices with different types of nonlinearities
\cite{Rumpf:2004,Khare:2006,Rumpf:2008,Johansson:2010,khare2013,Tieleman_2014,Mithun:2018},
Toda and Morse lattices \cite{Likhachev:2006}, etc. The localized wave
patterns that naturally emerge in such systems as a result of the
interplay between lattice dispersion and nonlinearity play a crucial
role
in the thermalization and lattice dynamics~\cite{Flach:1998}.

Our study is partially motivated by recent findings regarding the statistics of
different discrete nonlinear physical systems
\cite{Mithun:2019,Mithun:2018} in which, as a core mechanism, the
relaxation of nonlinear localized excitations and related ergodization
is considered. In the many-body systems, the ergodization demands
infinite-time averages of an observable during a microcanonical
evolution to match with their proper phase space averages.                                
In a class of dynamical systems where ergodization timescales
sensitively depend on the control parameters,
dynamical glass behavior is postulated to be a generic system
property on the route towards the integrable limits
\cite{Mithun:2019}. This glassy behavior is further attributed to the
short range network in the action space. Hence it is interesting to
explore the thermalization in a system where both an integrable a 
non-integrable (yet physically relevant) limits can exist as a suitable
parameter is varied. One such model is the Salerno
model (SM)~\cite{salerno1991discrete}.
On the one hand, and despite the two decades that have ensued since
the attempt to thermodynamically describe the discrete nonlinear
Schr{\"o}dinger (DNLS) model~\cite{Rasmussen:2000,kev2009}, the 
problem
continues to attract significant attention; see,
Refs.~\cite{gradenigo2019localization,PhysRevA.102.033301,PhysRevLett.122.084102}
for recent studies. On the other hand, the SM has provided an
excellent
platform for exploring the interplay between nonlinear localized structures
and near-linear extended ones, between nonlinearity and dispersion,
on the path between integrability and non-integrability; see,
e.g.,~\cite{boris}
for a recent review. 

Bearing the above features in mind, our aim here is to explore the
statistical mechanics and thermalization properties of the Salerno model.
In addition to interpolating between the fully integrable
Ablowitz-Ladik (AL) and the DNLS models
\cite{Johansson:2004,ablowitz1976nonlinear,salerno1991discrete,salerno1992quantum,melvin2009discrete},
the SM incorporates coexistence and competition of nonlinear
dispersion and nonlinear local interactions.
In the present work, we adopt a grand-canonical description of SM,
decomposing
the parameter space of energy and norm densities (corresponding to the
conserved quantities of the
energy and total norm, respectively) into Gibbs and non-Gibbs regimes.
In the former, we expect ``regular'' thermalization. In the latter
 non-Gibbs regime, 
 we expect to
 encounter
 energy localization in the form of long-living nonlinear excitations
 in line with the corresponding DNLS prediction
 of~\cite{Rasmussen:2000}.
One part of the intriguing story is that as the SM parameters are varied
homotopically interpolating between the DNLS and the AL models, we should be
able to observe that regions that used to belong in the non-Gibbs
regime will now ``regularly'' thermalize as we approach the AL
limit. That is,
we should be able to observe a key change in behavior parametrically
along the paths of parametric variation considered. However, there are
also
additional features that add to the complexity of the story.

While the general expectation previously was that the non-Gibbs regime
is non-ergodic, a recent study of the DNLS lattice using the
statistics of excursion times of equilibrium
Poincar{\'e} manifolds and finite time average distributions of an
observable has shown that a part of the non-Gibbs regime is weakly
ergodic~\cite{Mithun:2018, Danieli:2017, Mithun:2019}
in the setting of finite lattices. The non-ergodicity may be a feature
of the thermodynamic limit of infinite lattices. 
Motivated by this array of recent developments and challenging
observations,
we  numerically investigate the thermalization in the SM with respect to
both the Gibbs and
non-Gibbs regimes. We will utilize a combination of thermodynamic
tools,
including the transfer integral approach, and direct numerical
simulations,
and measures of both statistical properties (e.g., the probability
distribution function (PDF) of different amplitudes) and dynamical properties (such
as
Lyapunov exponents) in order to characterize the system, including for
different
lattice sizes, so as to address the fundamental question of the
behavior
of the SM under parametric, initial condition and lattice size
variations.

Our presentation is structured as follows. In section II, we present
the fundamentals of the model. In section III, we lay the theoretical foundations for the
statistical mechanical analysis of the SM. In section IV, we present
the corresponding numerical analysis (both through statistical and
dynamical
diagnostics) and finally in section V we summarize our findings, and
present
our conclusions, as well as some directions of future research. The
Appendix
presents details of our theoretical analysis.

\section{Model description}

The SM can be considered as a dicretization of the
continuous fully integrable nonlinear cubic Schr\"odinger
equation. Its  equations of
motion~\cite{salerno1991discrete,salerno1992quantum,melvin2009discrete},
upon suitable rescaling of the inter-site coupling, 
read:
\begin{eqnarray}
\begin{split}
\label{eq1}
i\frac{d\psi_n}{dt}&=(\psi_{n+1}-2\psi_n+\psi_{n-1})\\&
+
(\mu |\psi_n|^2)(\psi_{n+1}+\psi_{n-1})+\gamma |\psi_n|^2\psi_{n},
\end{split}
\end{eqnarray}
where $\psi_n$ is the complex wave function at site $n$, $\gamma=2(1-\mu)$ and $\mu \ge 0$. The parameter $\gamma$
represents the strength of local nonlinear interaction and $\mu$
represents the strength of nonlinear dispersion (inter-site nonlinearity).  
In the limit $\mu
\rightarrow 0$ the model reduces to the standard DNLS equation with on-site (local) cubic
nonlinearity. On the contrary, the limit $\gamma =0$ corresponds to the completely integrable AL model~\cite{ gomez2006solitons,Kivshar:1994,kev2009}. 
A simple transformation $\psi_n \rightarrow \psi_n e^{i2t}$ leads to
\begin{eqnarray}
\begin{split}
\label{eq1a}
i\frac{d\psi_n}{dt}&=
(1+\mu |\psi_n|^2)(\psi_{n+1}+\psi_{n-1}){+}\gamma |\psi_n|^2\psi_{n}.\label{equa1}
\end{split}
\end{eqnarray}
The full set of invariants of motion in the completely integrable, AL limit is considered in \cite{cassidy2010chaos}, while the nonintegrable DNLS limit is characterized by two integrals of motion.
Therefore, regardless of the limits,  Eq.~\eqref{equa1} can be characterized by two
conserved quantities: norm $\mathcal{A}$ and Hamiltonian $\mathcal{H}$ \cite{cai1994moving}:
\begin{equation}
\begin{split}
\mathcal{A}&=\sum_{n=1}^N \mathcal{A}_n,~~\mathcal{A}_n =\frac{1}{\mu} \ln{|1+\mu|\psi_n|^2|} \\
\mathcal{H}&=\sum_{n=1}^N\Big[-\frac{\gamma}{\mu} \mathcal{A}_n+\psi_n\psi_{n+1}^*+\psi_n^*\psi_{n+1}+\frac{\gamma}{\mu} |\psi_n|^2 \Big],
\label{eq3}
\end{split}
\end{equation}
where $N$ is the total number of lattice nodes
and periodic boundary conditions are used. 

The SM equations (Eqs. \eqref{equa1}) can be derived from the Hamiltonian $\mathcal{H}$
\begin{equation}
\frac{d\psi_n}{dt}=\{\mathcal{H},\psi_n\}.
\end{equation}
with respect to the canonically conjugated pairs of variables $\psi_n$ and $i\psi_n^*$ defining the deformed Poisson brackets \cite{Cai:1994}
\begin{equation}
\label{eq4}
\{\psi_n,\psi_m^*\}=i(1+\mu|\psi_n|^2)\delta_{nm},~~
\{\psi_n,\psi_m\}=\{\psi_n^*,\psi_m^*\}=0.
\end{equation}

\section{Statistical Mechanics of the  Salerno network}

Here we attempt to clarify the thermalization properties of the SM
starting from the DNLS limit $\mu=0$, in
which the thermalization and statistical properties are extensively investigated \cite{Cai:1994,jackson1978nonlinearity,Rasmussen:2000,Iubini:2012, Mithun:2018}. After a brief remark on findings in the DNLS limit we probe the extension of the Gibbs approach to the SM with competing local and nonlocal nonlinearities. 

Applying the canonical transformation 
$\psi_n=\sqrt{A_n}\exp{(i\phi_n)}$, where $A_n$ and $\phi_n$ denote the
amplitude and phase, we obtain from Eq. (\ref{eq3}) the following
expressions for the conserved quantities:
\begin{equation}
\begin{split}
\mathcal{A}&=\frac{1}{\mu}\sum_{n=1}^N \ln{|1+\mu A_n|}, \\
\mathcal{H}&=\sum_{n=1}^N \big[-\frac{\gamma}{\mu^2} \ln|1+\mu \,  A_n| +\frac{\gamma}{\mu} A_n\\
& +\sqrt{A_n\,A_{n+1}}\cos{(\phi_n-\phi_{n+1})} ]  \big].
\label{eq33}
\end{split}
\end{equation}
The corresponding grand-canonical partition function of the SM can then be presented in a form
\begin{equation}
\textit{Z}= \int_{0}^{\infty} \int_{0}^{2\pi} \prod_{n=1}^N d\phi_n dA_n e^{-\beta (\mathcal{H}+\alpha \mathcal{A})},
\label{eq:Part_function}
\end{equation}
where parameters $\alpha$ and $\beta$ are introduced in analogy
with the chemical potential and the inverse
temperature~\cite{Rasmussen:2000} (i.e., they are the corresponding
Lagrange multipliers). This expression can be reduced to the integral form after the integration over the phase variables $\phi_n$:
\begin{equation}
\begin{split}
Z&=(2\pi)^N\int_{0}^{\infty}\prod_{n=1}^{N} dA_n \\& \times
I_0(2\beta\sqrt{A_nA_{n+1}})\,e^{-\beta \sum_m
	((-\frac{\gamma}{\mu^2}+\frac{\alpha}{\mu}) \ln|1+\mu \, A_n|
	+\frac{\gamma}{\mu} A_n) } ,
\label{a4}
\end{split}
\end{equation}
where $I_0$ stands for the modified Bessel function of the first kind (with index $0$).

In the thermodynamic limit of large systems $N\rightarrow \infty$,
the integral can be evaluated exactly using the eigenfunctions and
eigenvalues of the transfer integral operator (TIO) (Appendix
\ref{AP:SA}). From the
latter calculation, in the infinite temperature limit $\beta\rightarrow 0$, the following relation between the energy ($h=\mathcal{H}/N$) and norm ($a=\mathcal{A}/N$) density  can be derived
\begin{eqnarray}
h=  \frac{2(1-\mu) a^2}{( 1- a \mu )}, \label{anb6}
\end{eqnarray}
which for $\mu=0$ reduces to the corresponding relation of the DNLS lattice \cite{Rasmussen:2000, Johansson:2004, Mithun:2018}. {In terms of the largest eigenvalue $\lambda_0$ of the kernel Eq.~\eqref{eq:kernal}, the norm density, $a$ can be expressed as $a=\int_0^{\infty} y_0^2(A) A dA$, where $y_0^2(A)$ represents the probability distribution of amplitudes $P(A)$ corresponding to the largest eigenvalue}.

Following the statistical mechanical analysis of the Appendix, we
distinguish the Gibbs regime in the $(a,h)$ parameter space by
determining the characteristic phase curves $\beta\rightarrow 0$  and
$\beta\rightarrow \infty$. While formally the first one separates the
microcanonically inaccessible regime from
the Gibbs region of phase space, the second one separates regions
characterized by positive temperature ($\frac{1}{\beta}>0$) from those with negative
temperature ($\frac{1}{\beta}<0$) whose accessibility is experimentally and
numerically proven
and which leads to the prolonged emergence of coherent structures.

The transition curve $\beta\rightarrow \infty$ can be determined from 
\begin{equation}
\begin{split}
h&= -\frac{\gamma}{\mu} a+\frac{\gamma}{\mu^2} (e^{\mu a} -1)-\frac{2}{\mu}(e^{\mu a}-1),~~\text{and}~~ \\ 
a& =\frac{1}{\mu}\ln(1+\mu d).
\label{eq5}
\end{split}
\end{equation}
by minimizing the Hamiltonian Eq. (\ref{eq3}) with the plane-wave solution in a form $\psi_n =\sqrt{d} e^{in\theta}$ and taking $\theta=\pi$.

To clarify the thermalization properties we calculate the amplitude
probability density function ($P(A)$) and the excursion time probability ($P_{+})$.
The $P(A)$
obtained from a direct numerical simulation of Eq.~\eqref{eq1} is compared with the one calculated via the TIO approach~\cite{Rasmussen:2000}. 
On the other hand, the time intervals which the local norms spend between two consecutive intersections of the plane $A_n=a$ (the Poincar{\'e} section) form the excursion time distribution $P_{+}(\tau)$, where $+$ denotes $A_n>a$ and $\tau_n(i)=t_n^{i+1}-t_n^i$. The distribution has the average $\mu_{\tau}$ and the standard deviation $\sigma_{\tau}$.  The value of $\sigma_{\tau}$ can be associated with the divergence of the average excursion time and weak nonergodicity in the lattice~\cite{Mithun:2018,Danieli:2017}.  A third measure of thermalization is the finite time average (FTA) of the observable,
\begin{equation}
 A_{n,T}=\frac{1}{T}\int_0^TA_n(t)dt.
\end{equation}
 The distribution of the FTA for a set of trajectories is characterized by the first moment $m_1(T)$ and the second moment $m_2(T)$. For an ergodic regime, at large time $m_2(T \rightarrow \infty) \rightarrow 0$ \cite{Mithun:2019,Danieli:2019}.

As an additional diagnostic, we estimate the maximal Lyapunov
Characteristic Exponent (mLCE) $\Lambda$ \cite{benettin:1980a,benettin:1980b,S2010} which is in general a measure
of the degree of chaos in the system.
More concretely, we derive the evolution equations for small
perturbations $\chi_n(0)$ of the initially injected plane-wave profile $\psi_n(0)$
adopting the standard procedure based on the linearization in the
presence
of small perturbations and numerically solve the obtained variational equations
\begin{equation}
\begin{split}
i\frac{d \chi_n}{dt}&= (1+\mu|\psi_n|^2)(\chi_{n+1}+\chi_{n-1})+\mu (\psi_{n+1}+\psi_{n-1})\\& \times (\psi_n^{\ast}  \chi_n+\psi_n \chi_n^{\ast})+\gamma (2 |\psi_n|^2\chi_n+ \psi_n^2 \chi_n^{\ast}).
\end{split}
\label{eq:mlce}
\end{equation}
The mLCE is then obtained as 
\begin{equation}
\begin{split}
\Lambda= \lim_{t \rightarrow \infty} \lambda(t) =\lim_{t \rightarrow \infty} \frac{1}{t}\ln{\frac{||\pmb{\chi}(t)||}{||\pmb{\chi}(0)||}},
\end{split}
\label{eq:mlce1}
\end{equation}
with $\lambda(t)$ denoting the so-called finite time mLCE (ftmLCE), $\pmb{\chi}(t)=(\chi_1,\chi_2,...,\chi_N)$ being the deviation vector 
and $||\cdot ||$  the usual Euclidean  norm.

 \begin{figure}[!htbp]
 	\includegraphics[angle=0, width=\linewidth]{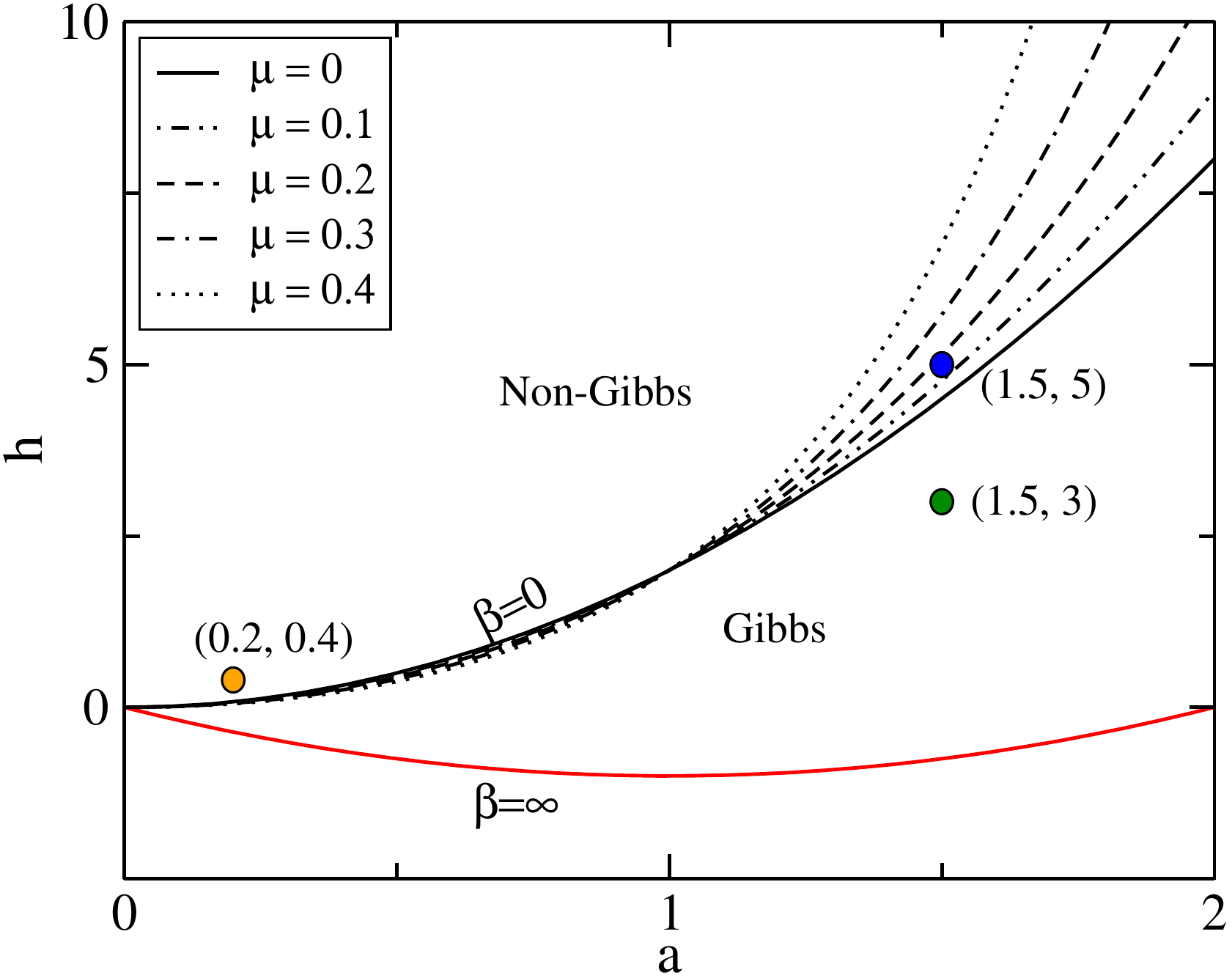}
 	\caption{The norm and energy density parameters space $(a, h)$ of the SM. The area between the curves $\beta=0$ (black) and
          $\beta\rightarrow \infty$ (red) 
         in the parameter space
          $(a,h)$ denotes the Gibbs regime of anticipated thermalization
          (Eq.~\eqref{eq:Part_function} is valid) for a few values of
          the parameter $\mu$ ($=$ 0,\, 0.1,\, 0.2,\, 0.3,\, 0.4). The area above
          $\beta= 0$ will be referred to as non-Gibbs, where Eq.~\eqref{eq:Part_function} is invalid.  The region below $\beta\rightarrow \infty$ is forbidden for any microcanonical states. The green, blue and orange symbols respectively represent ($a=1.5$, $h=3$),  ($a=1.5$, $h=5$) and ($a=0.2$,$h=0.4$).} \label{fig:phase}
 \end{figure}

In order to investigate thermalization in the SM we perform 
numerical experiments and base our corresponding analysis on the 
system's phase diagram illustrated in 
Fig.~\ref{fig:phase}, in the parameter space $(a,h)$. The red curve
corresponds to the zero temperature limit $\beta\rightarrow \infty$,
while black curves correspond to the infinite temperature limit,
$\beta\rightarrow 0$ for a few values of parameter $\mu$ in the interval
$0$ to $0.5$. 
These are based on the analytical predictions of these limits given
above
(and derived in the Appendix).
The region between $\beta=\infty$ and $\beta=0$ lines denotes the
Gibbs regime of the SM where the approach based on the
grand-canonical-Gibbs statistics (Eq.~\eqref{eq:Part_function}) is
applicable \cite{Rasmussen:2000, Mithun:2018}) and thus the model
is expected to thermalize. Outside this region, a non-Gibbs regime
featuring the presence of coherent structures is identified where,
however,
for finite domains only weak non-ergodicity may be present as discussed
in~\cite{Mithun:2018}. We now proceed to analyze our numerical computations at
different selected points within this parameter space bearing in mind that a key
feature of the SM case is that regimes identified as non-Gibbsian for the DNLS
model of $\mu=0$ can be Gibbsian for larger values of $\mu$. 

 \begin{figure}[!htbp]
 	\includegraphics[angle=0, width=\linewidth]{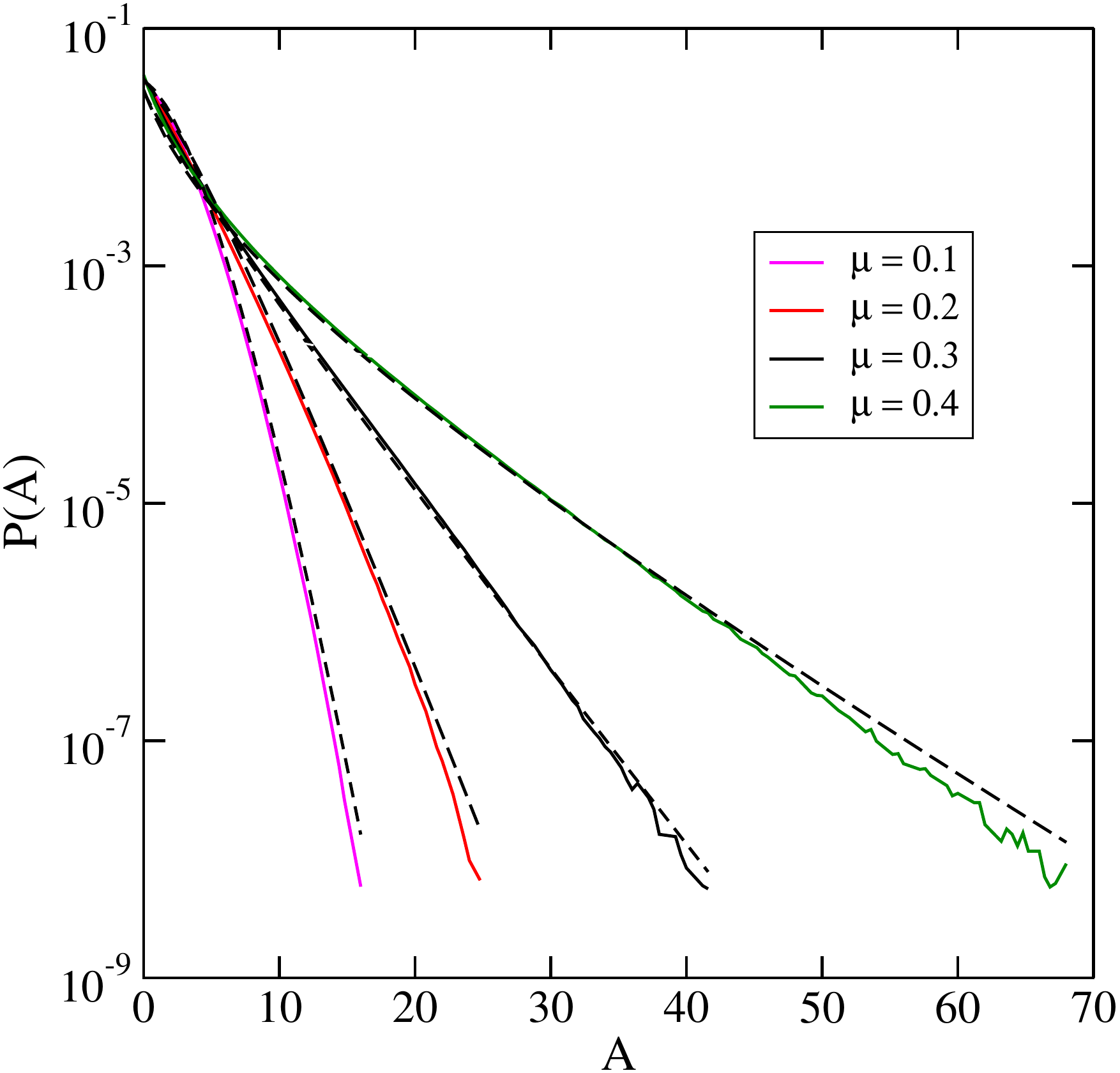}
 	\caption{The PDF of local norms (amplitudes) $A_n =|\psi_n|^2$ for different values of $\mu$  for the parameter set ($a=1.5$, $h=3$).  Dashed curves represent the results obtained analytically by the TIO method, while solid curves show the numerically generated PDFs. The total integration time is $T =10^7$. TIO curves nicely fit to numerical ones in the Gibbs regime. 
 	}
 	 \label{fig:pda}
 \end{figure}
 \begin{figure*}[!htbp]
	\includegraphics[angle=0, width=\linewidth]{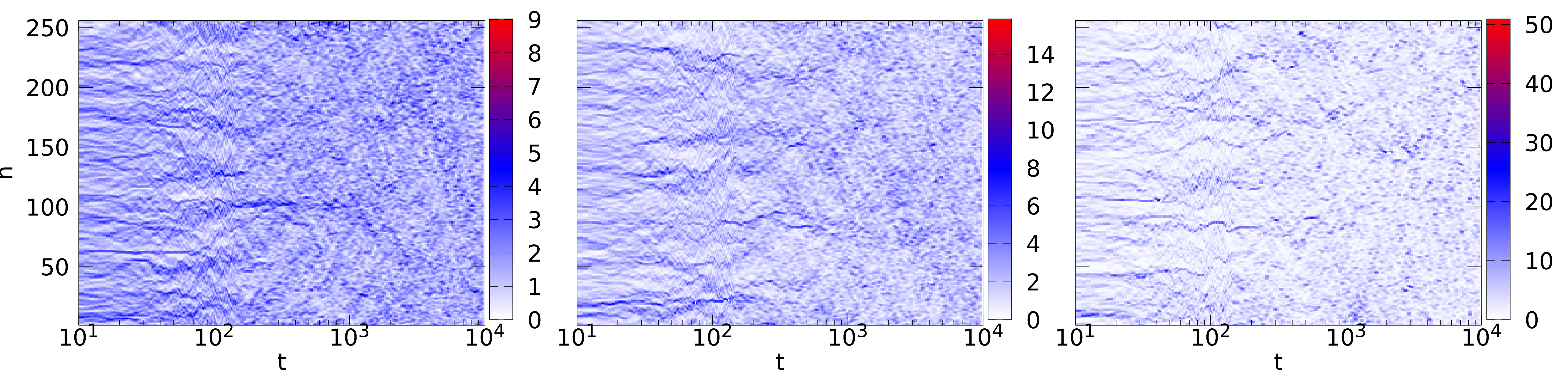}
		\caption{
		The amplitude profiles in the Gibbs regime for three different $\mu$ values (a) $\mu=0.05$, (b) $\mu=0.2$, and (c) $\mu=0.5$) at ($a=1.5,\, h=3$) for the system size $N=256$. 
		} 
	\label{fig:breathergibbs}
\end{figure*}
  \begin{figure}[!htbp]
  	\includegraphics[angle=0, width=\linewidth]{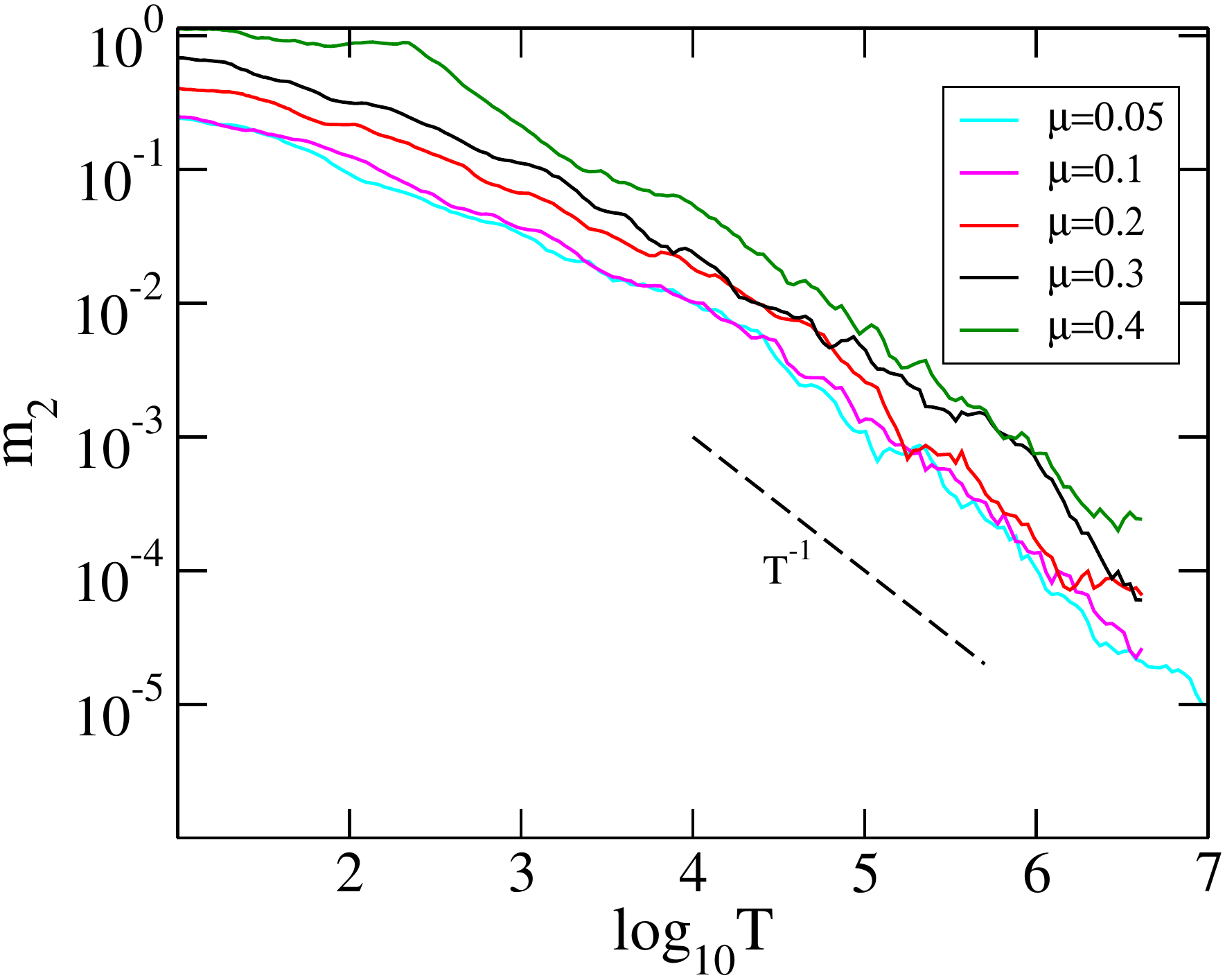}
  	\caption{The evolution of the second moment ($m_2(T)$) of FTA distributions of the integral local norms for the different values of $\mu$ corresponds to ($a=1.5$, $h=3$).  The dashed line represents $m_2\propto T^{-1}$.   
  	} 
  	\label{fig:ftaa}
  \end{figure}
\begin{figure}[!htbp]
	\includegraphics[angle=0, width=\linewidth]{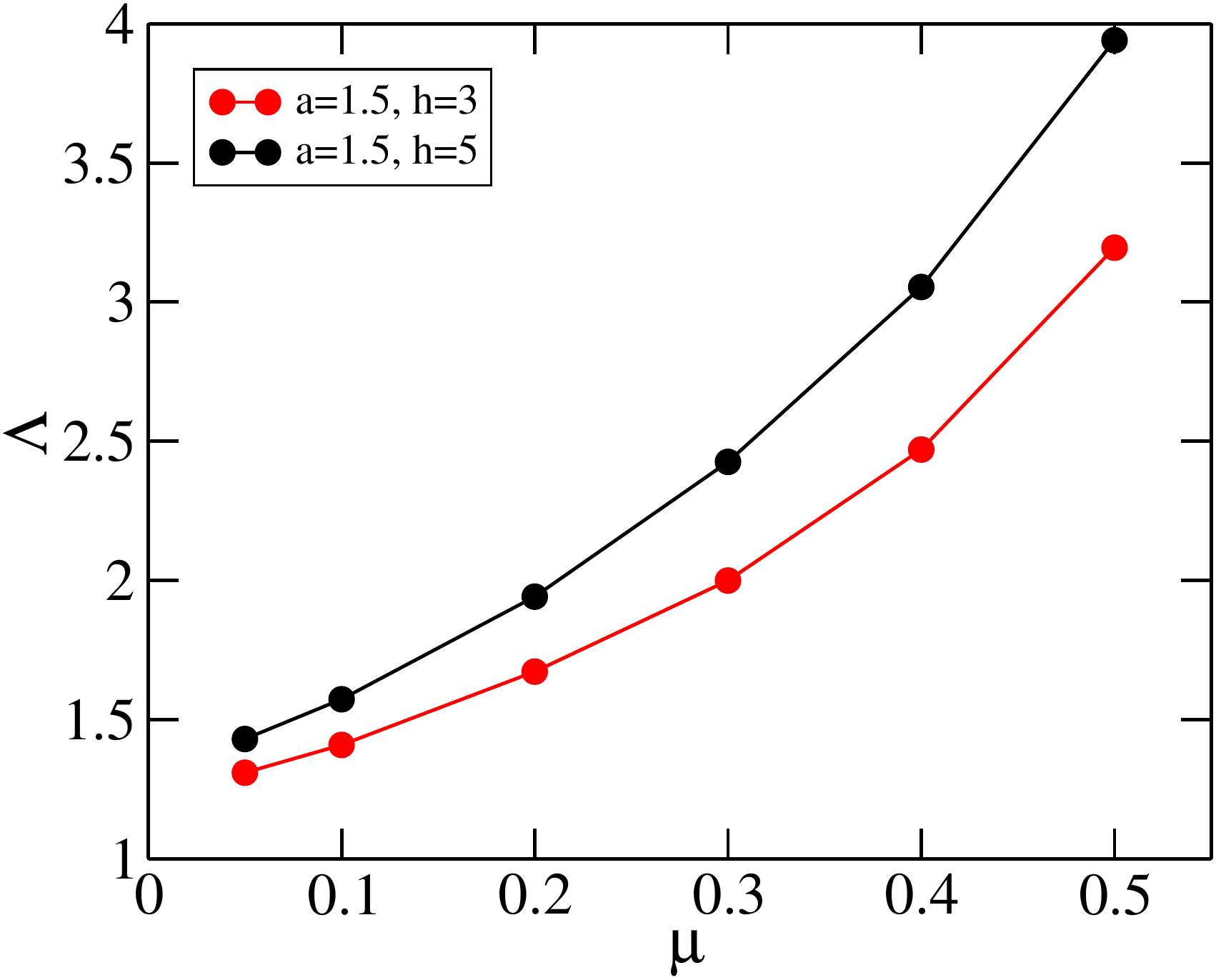}
	\caption{The mLCE ($\Lambda$) for ($a=1.5,\, h=3$) (red curve)  and ($a=1.5,\, h=5$) (black curve) as a function of $\mu$ for $N=256$. In both cases the value of $\Lambda$ increases with $\mu$. See the Appendix \ref{lyp_3} for the finite time mLCE $\lambda(t)$.
	} 
	\label{fig:lyp}
\end{figure}
 \section{Numerical Analysis}
 In this section, we describe the numerically obtained results for various $\mu$ values corresponding to both the Gibbs and non-Gibbs regimes. This section aims to explore the thermalization properties in these regimes induced by the competing nonlinearities, local nonlinear interaction and nonlinear hopping, which cause self-trapping and  nonlinear  dispersion, respectively.  
 We solve Eq.~\eqref{eq1} numerically by using an explicit Runge-Kutta algorithm of order 8, called DOP853 \cite{freelyDOP853,hairer1993solving,mine-01-03-447}. We set the  relative energy error $\left| \frac{\mathcal{H}(t)-\mathcal{H}(0)}{\mathcal{H}(0)} \right|$ and norm $\left| \frac{\mathcal{A}(t)-\mathcal{A}(0)}{\mathcal{A}(0)} \right|$ error threshold $10^{-4}$.
 Initially a small complex random perturbation is added to the plane waves 
$$\psi_n(0)=\sqrt{d}\exp{(i\phi_n)}, $$ 
where, $d=\mu^{-1}(\exp(a\mu)-1)=\mathcal{A}/N$. 
Unless otherwise mentioned we use a total integration time of $T=10^7$
and a system size of $N=256$.

 \subsection{Gibbs regime}
 We consider the parameter set ($a=1.5,\,h=3$) which is in the Gibbs
 regime irrespective of the $\mu$ value considered (see
 Fig.~\ref{fig:phase}). The numerically calculated PDFs, $P(A)$
 (solid curves in Fig.~\ref{fig:pda}) show that the amplitude $A$
 increases with the increase of $\mu$.
 In order to ensure that the obtained $P(A)$ represents a thermalized
 state, we compare them with the probability distributions obtained by
 the transfer integral operator (TIO) approach
based on the corresponding dominant (squared) eigenvector of the TIO approach
(see, e.g., Appendix I and also~\cite{Rasmussen:2000}).
Our results indicate that TIO solutions (dashed curves in
Fig.~\ref{fig:pda}) and numerical results match very well,
which corroborates the anticipated thermalization in this regime.  
 
 The  amplitude profiles corresponding to the case ($a=1.5,\,h=3$) for
 three different $\mu$ values are shown in
 Fig.~\ref{fig:breathergibbs}. Though high amplitude nonlinear
 localized excitations emerge in the system as a result of the
 modulational
 instability of the initial condition, they are all rather short lived
 and the long time evolution of the system appears to be thermalized
 into a phononic bath and without evidence of any kind of persistent
 localization.
 The thermalization is further verified from the second moment,
 $m_2(T)$ of the FTA of local integral norms (Fig.~\ref{fig:ftaa}),
 which decays as $1/T$ at large times for all values of the considered
 $\mu$. 
 Additionally, the calculation of the mLCE shows that $\Lambda>0$ (red points in Fig.~\ref{fig:lyp})
 which is a signature of chaoticity in the system. Interestingly,
 we find that the latter grows exponentially  as a function of $\mu$.

    \subsection{Non-Gibbs regime}
   
\begin{figure}[!htbp]
 	\includegraphics[angle=0, width=\linewidth]{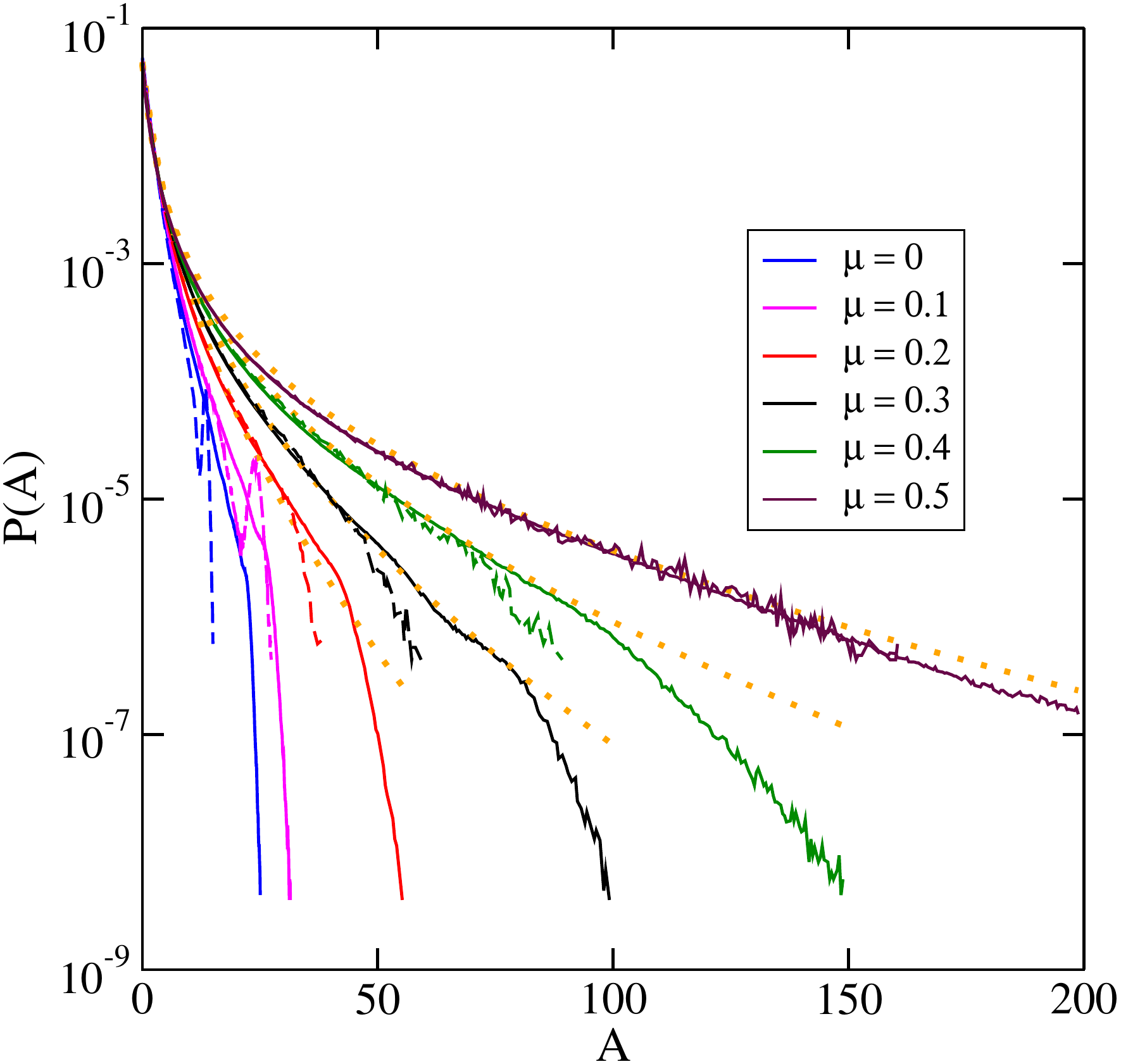}
 	\caption{The PDF of local norms
          (amplitudes) $A_n =|\psi_n|^2$ for different values of
          $\mu$ for the parameter set  ($a=1.5$, $h=5$). This arrangement is in the Gibbs regime
          for $\mu>0.17$ and otherwise in the non-Gibbs regime. Different
          curves are associated to $\mu=0,\, 0.1,\, 0.2,\, 0.3,\,
          0.4,\, 0.5$ as shown in the legend. {Dotted orange curves represent
          the results obtained analytically by the TIO method, while
          dashed and solid lines show the numerically generated $P(A)$ at two different times $T =10^5$ and $T =10^7$, respectively. TIO
          curves fit 
           the numerical ones in the Gibbs regime, except for the large amplitude tail region. }}

 	 \label{fig:pdab}
 \end{figure}

To investigate the thermalization in the non-Gibbs regime, we consider
the parameter set ($a=1.5,\,h=5$), the blue {colored} point in
Fig.~\ref{fig:phase}. For this parameter set, the critical value
$\mu_c\approx 0.17$ sets the transition point from Gibbs to non-Gibbs
regime. That is, for $\mu>\mu_c$, ($a=1.5,\,h=5$) is in the Gibbs
  regime, where TIO solutions match exactly with the numerically
  calculated $P(A)$ as
  shown in Fig.~\ref{fig:pdab}.
 \begin{figure*}[!htbp]
	\includegraphics[angle=0, width=\linewidth]{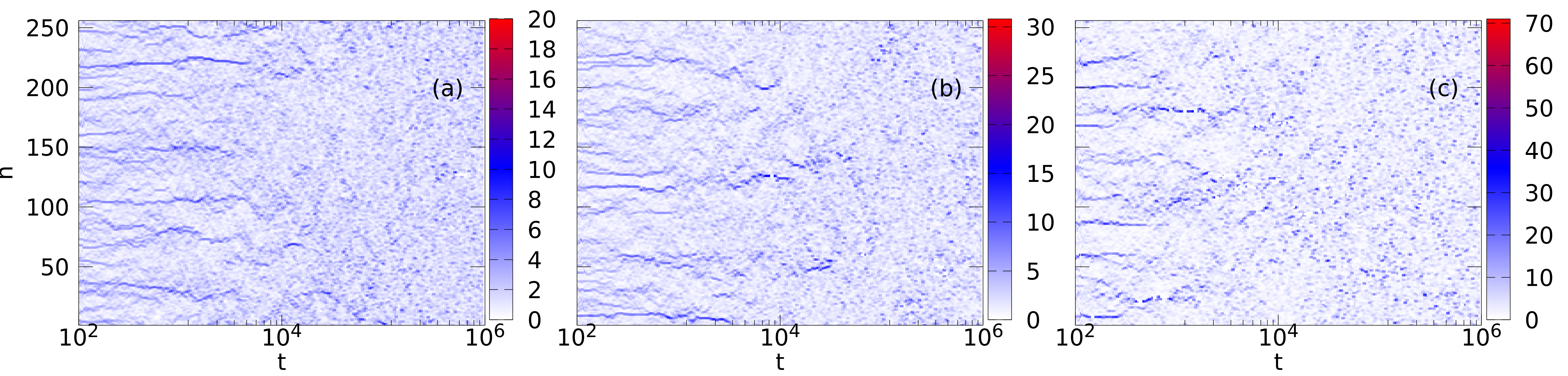}
		\caption{
                  In panels (a), (b) and (c),
                  the amplitude profiles are respectively shown in the non-Gibbs regime
                ($\mu=0.05$),
                near the critical point ($\mu=0.2$) and in the Gibbs
                regime ($\mu=0.5$) for $a=1.5,\, h=5$ and $N=256$. In
                the non-Gibbs phase (panel (a)), the appearance of localized
                breathing structures is significantly more prominent
                than in the
                Gibbs regime (panel (c)).}
   	\label{fig:breathernongibbs}
\end{figure*}
\begin{figure}[!htbp]
 	 	\includegraphics[angle=0, width=\linewidth]{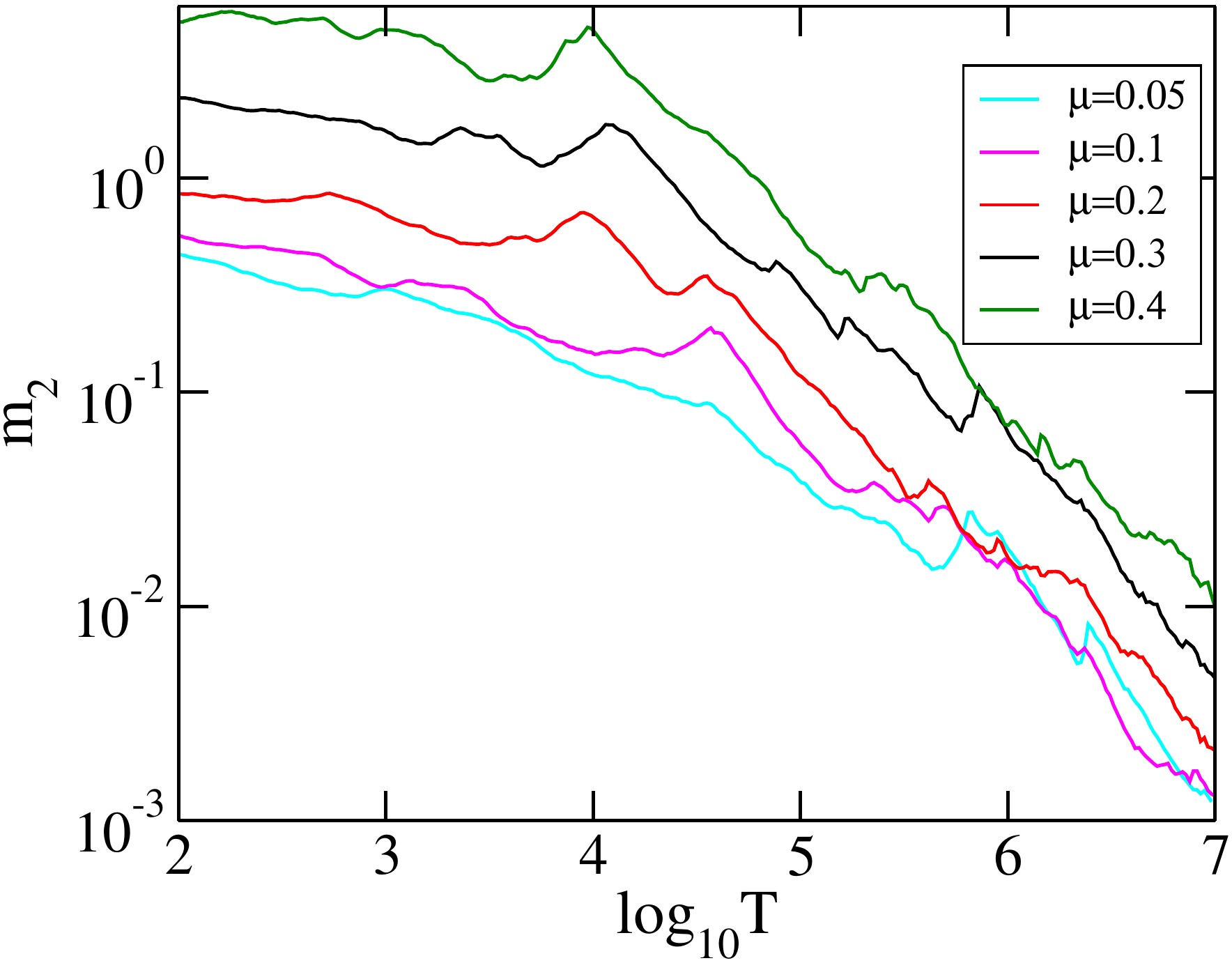}
 	\caption{
 	The second moment ($m_2({T})$) of FTA distributions versus
          $\mu$ for ($a=1.5,\, h=5$). The $\mu<\mu_c (\approx 0.17$) case represents the non-Gibbs regime.
 	} 
 	 \label{fig:ftab}
 \end{figure}
\begin{figure}[!htbp]
	\includegraphics[angle=0, width=\linewidth]{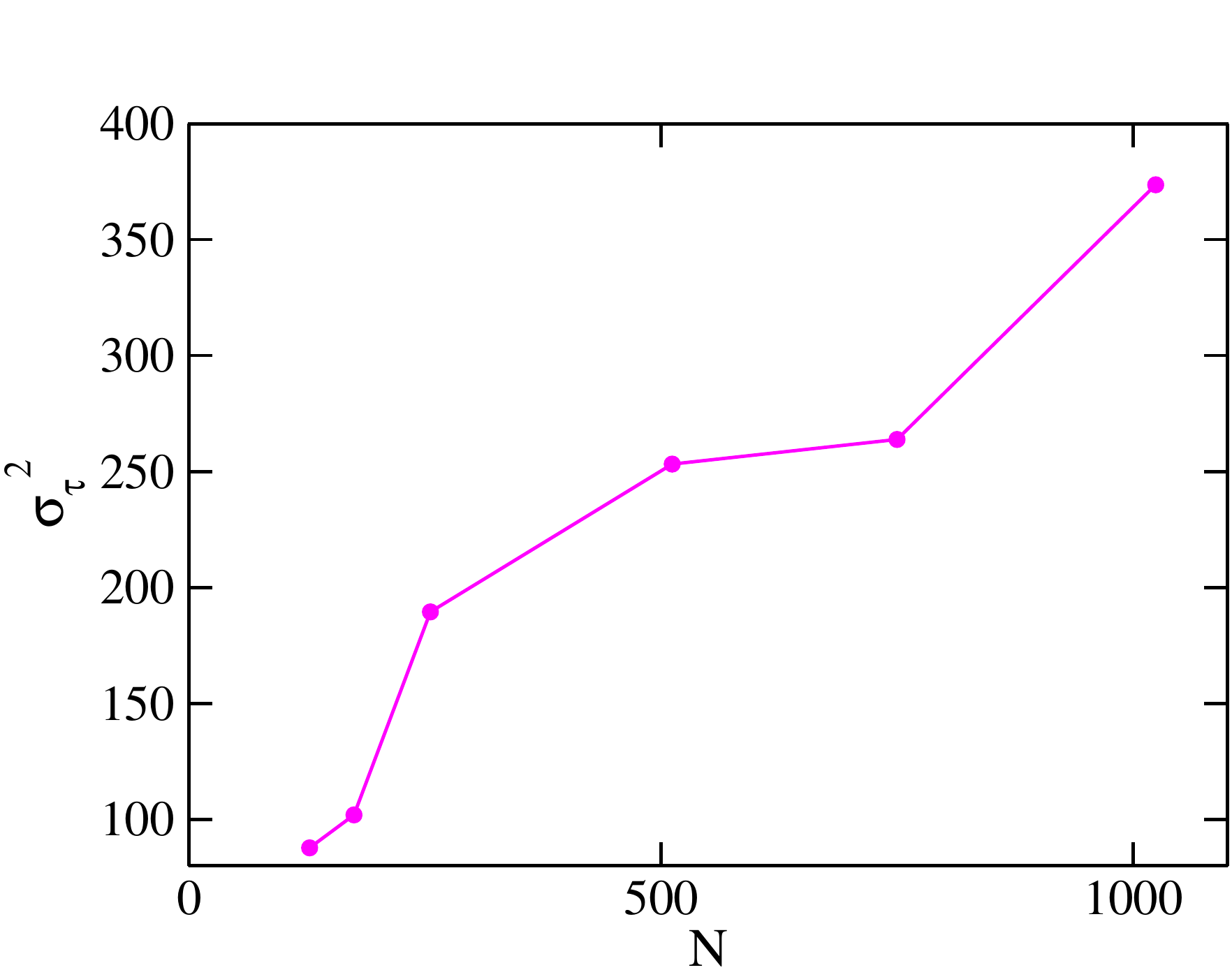}
	\caption{The variance $\sigma_{\tau}^2$ for $\mu=0.05$ and ($a=1.5$, $h=5$) as a function of system size $N$. This case corresponds to the non-Gibbs regime. The total integration time $T=10^7$. The results are averaged over 5 different initial conditions. 
	} 
	\label{fig:life}
\end{figure}
 \begin{figure*}[!htbp]
	\includegraphics[angle=0, width=\linewidth]{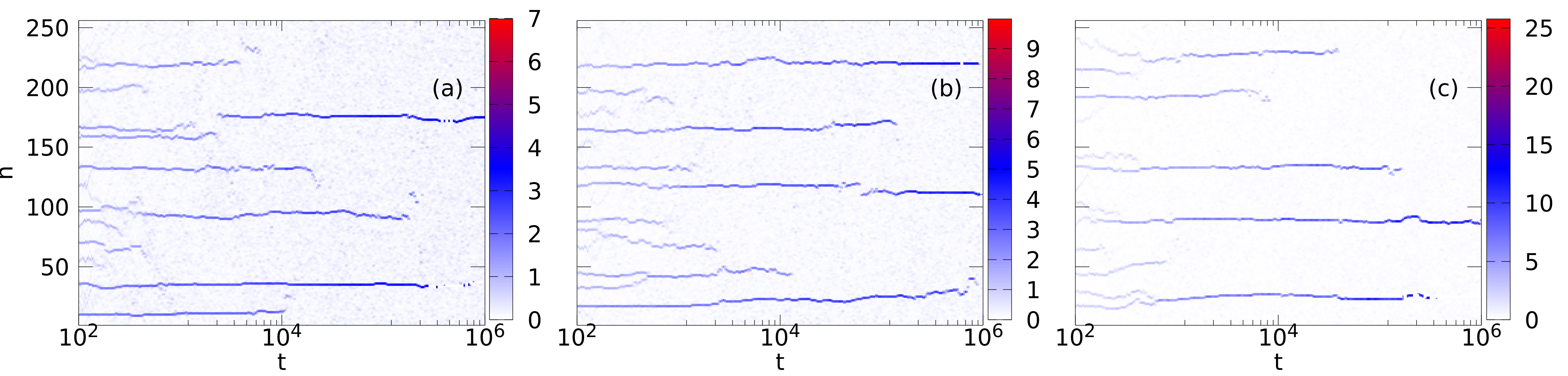}
		\caption{
		The amplitude profiles in the non-Gibbs regime for (a) $\mu=0.05$, (b) $\mu=0.2$ and (c) $\mu=0.5$ at $a=0.2,\, h=0.4$ for $N=256$. } 
	\label{fig:breathernongibbsd}
\end{figure*}

{For $\mu<\mu_c$, the tail of $P(A)$ develops a bump (at
suitably
large amplitudes) at initial times, which can be associated with the accumulation of
large amplitude nonlinear excitations. In spite of it, we observed an
exponential cutoff that might be related to the finite size of the
system, which has been shown to affect all
statistical measures we calculated in the non-Gibbs regime.
It is worthwhile to observe that this bump is no longer present in the
cases of $\mu>\mu_c$ (while the cutoff is still featured) and at large times for $\mu<\mu_c$}.
To corroborate this observation in the system, but also to explore
the role of the finite size effects, we first plot the amplitude
profiles as a function  of time for three different values of $\mu$,
Fig.~\ref{fig:breathernongibbs}.
Interestingly, for all values of $\mu$, both those that belong to the
Gibbs regime
($\mu=0.5$ and $\mu=0.2$ of panels (c) and (b); the latter is near and
above the critical value) and even in the non-Gibbs regime
($\mu=0.05$), the decay of initially generated large amplitude
structures is eventually observed.
Nevertheless,
the coherent structures are significantly more prominent in the
non-Gibbs case.
The latter case of panel (a) represents the possibility of a
non-Gibbsian
regime which, however, within a finite lattice manifests the features
of quasi-ergodic behavior as has been discussed in~\cite{Mithun:2018}.
The second moment of the FTA of the integral local
norms, shown in Fig.~\ref{fig:ftab}, decays over time for the considered $\mu$ values. 
The curves, however, do not indicate any distinctive behavior that differentiates the dynamics in Gibbs ($\mu > \mu_c$) and non-Gibbs regimes ($\mu < \mu_c$). A similar result (in the sense of not distinguishing between
the Gibbs and non-Gibbs regimes) is also found  for the mLCEs in Fig.~\ref{fig:lyp}.

 In line with recent developments in the
field~\cite{Mithun:2018,Mithun:2019},
it is quite relevant to explore the effects of the finite size of our
computation
and their implications in connection with the infinite domain
thermodynamic
analysis, e.g., implicit in the TIO calculations. 
To explore the role of finite system size, we calculate the variance,
$\sigma_{\tau}^2$ of the excursion time distribution for different
system sizes as depicted in Fig.~\ref{fig:life}.  Upon increasing
the system size $N$, the variance $\sigma_{\tau}^2$ increases
indicating the significance of the finite size effect in the
thermalization of the non-Gibbs regime for $h \rightarrow h _{\beta =
  0}$. The somewhat non-smooth nature of the growth might be related
to the small
number of initial conditions used for the averaging.  
{Nevertheless, it can be conjectured that in the limit $N \rightarrow \infty $ the excitations in the non-Gibbs regime will be persistent}. 
It is also
worthwhile
to note that in the calculation of the variance, $\sigma_{\tau}^2$, the
excitations whose life time is higher than the total integration time,
$T=10^7$, are not considered.
On the other hand, the decay of the
second moment, $m_2$ of the FTA distribution shown in Fig.~\ref{fig:ftab}
indicates that there are no such long-living excitations. Since we observed that the finite size plays a crucial  role in the non-Gibbs regime for $h \rightarrow h _{\beta =
  0}$, we next consider a point in the parameter space $(a,h)$, that is far from $h _{\beta =
  0}$. This higher ratio of $\frac{h}{h _{\beta= 0}}$ can be 
  more straightforwardly obtained for a small norm and we fix ($a=0.2$, $h=0.4)$ (orange symbol in Fig.~\ref{fig:phase}). As the ratio $\frac{h}{h _{\beta= 0}}$ becomes larger, the contribution of the nonlinear interaction term in Eq.~\eqref{eq3} is higher due to the boundedness of the kinetic energy \cite{Rasmussen:2000}.
The  corresponding amplitude evolutions are shown in
Fig.~\ref{fig:breathernongibbsd} for three different values of
$\mu$. For all these $\mu$ values, the parameter set ($a=0.2$,
$h=0.4)$ is in the non-Gibbs regime and accordingly, the amplitude
evolution shows the presence of at least one excitation with life time
greater than the total computation
time. This hints at a non-ergodic behavior for the considered long but finite time.  

In summary, our analysis has illustrated that the Gibbs regime for the
DNLS model remains Gibbsian for the SM of the present considerations
in line with the corresponding analytical results. The situation
becomes considerably more interesting beyond the thermalization limit
of $\beta=0$ for the DNLS;
progressively the thermalization region of the $(a,h)$
space
expands upon increase of $\mu$ rendering non-thermalized parameter
ranges for smaller $\mu$ thermalized as $\mu$ grows past a certain
threshold. However, there exist additional features to consider. On
the one hand, the finiteness of the lattice plays a considerable role
as
to whether non-ergodicity will be preserved and it is indeed found
that
finite lattices may lead to an apparent ergodization. Nevertheless, as
the domain size grows to infinity, so does the life time of the
high-amplitude,
nonlinear excitations in line with the thermodynamic model prediction.
At the same time, the thermalization in the non-Gibbs regime heavily
depends on the $\frac{h}{h _{\beta= 0}}$ ratio.
When the latter becomes larger, then the system may find itself
in a non-ergodic regime even in the case of the finite lattice
(long-time)
computations. Admittedly, these features are the ones that are
qualitatively
reported here and merit additional quantification through extensive
and highly-demanding (in their computation time and accuracy)
computations.
Nevertheless, we trust that the above results offer useful insights
in this direction.

 \section{Conclusions \& Future Challenges}

The thermalization in the nonintegrable regime of the Salerno model (SM) has been
investigated by using a combination of analytical and numerical techniques.
More specifically, we have complemented the
transfer integral operator (TIO) analysis by performing relevant long-time numerical
simulations. In the latter,
we have used a set of
diagnostics such as the  probability distribution of amplitudes and
finite time averages of local probability density (and its moments)
as well as, e.g., Lyapunov characteristic exponents.
A key feature of the model is the coexistence of  local nonlinearity
and nonlinear dispersion. The competition between these two  effects
sets two limits, a nonintegrable and an
integrable one
in the system, namely the discrete
  nonlinear Schr{\"o}dinger (DNLS) and the Ablowitz-Ladik (AL) models.
One can then mono-parametrically interpolate between
these limits. Our study has mainly focused on the non-integrable
limit, but
providing a sense of the qualitative variation of the thermodynamic
properties
as the integrable limit is gradually approached.  In the former limit
the statistical mechanics of the system yields a phase diagram in the
parameter space of energy and norm densities. An infinite temperature
line decomposes the phase space into Gibbs and non-Gibbs regimes.
The freedom afforded by the Salerno model is that
we can achieve a transition from a non-Gibbs to a Gibbs regime for the same norm
and energy parameters, upon variation of 
the strength of the nonlinear dispersive term ($\mu$) interpolating
between the
two limits. Our analysis has shown that in the Gibbs regime, the SM is
ergodic, and that as $\mu$ is increased towards the integrable limit,
the region of the two-parameter space that thermalizes progressively expands.
Nevertheless, there are some additional important features.
More specifically, in the non-Gibbs regime, the ergodic properties
heavily depend on the initial conditions and the ratio $h/{h_{\beta
    =0}}$. Additionally, the finite system
size plays a crucial role and the TIO predictions are (expected to be) 
genuinely valid
in the thermodynamic limit.

We believe that  the present work provides insights into the thermalization of lattice
systems
and especially as we start approaching the integrable limit, including
the role of parameters such as the lattice size and how ``deep''
in the non-Gibbs regime the initial conditions are. However, our results 
also
raise a number of significant questions for future studies.
Specifically, it is important to understand to what degree
we can extend the present picture further towards the integrable limit and
what happens in its immediate vicinity. Here, an important issue that arises
is that additional conservation laws come into
play~\cite{cassidy2010chaos}
and how these can be
incorporated in thermodynamic considerations. The role
they play in modifying (or dynamically constraining) the picture
is something that is especially relevant to understand, in the
immediate
vicinity of the integrable limit and then further away from it.
{ In that vein, revisiting also related studies exploring the
creation and disappearance, as well as mobility of discrete breathers
as they interact with the phonon bath~\cite{Rasmussen:2000:a} (and how these mechanisms
change while approaching integrability) would be an especially interesting direction.}
Additionally, while we have illustrated the relevance of finite size
and of ratios such as $h/{h_{\beta
    =0}}$, obtaining a quantitative characterization of their role
and of, e.g., the scaling
dependence on the coherent structure (average) lifetime on them
emerges as an especially relevant problem. This would greatly help
appreciate the influence of quasi-ergodicity ideas such as those
put forth in~\cite{Mithun:2018,Mithun:2019}. Lastly, all of the above
features have been explored in one-dimensional contexts yet it would
be rather natural to extend considerations to higher-dimensional
one. These topics are presently under
consideration
and findings will be presented in future publications.


\section{Acknowledgements}
This material is based upon work supported by the US National  Science  Foundation  under  Grant DMS-1809074 (PGK). A.M.~acknowledges support by Ministry of Education, Science and Technological Development of the Republic of Serbia grant 451-03-68/2020-14/200017.  AK would like to thank Indian National Science Academy (INSA) for
the award of INSA Senior Scientist position at Physics Department, Savitribai Phule Pune University, Pune, India. B.M.M.~and Ch.S.~were supported by the National Research Foundation (NRF) of South Africa and thank the  High Performance Computing facility of the University of Cape Town, as well as the Centre for High Performance Computing (CHPC) of South Africa for providing their computational resources. 
The work at Los Alamos National Laboratory was carried out under the auspices of the U.S. DOE and NNSA under Contract No. DEAC52-06NA25396

\appendix

\section{The grand-canonical approach and the nonintegrable
  SM} \label{AP:SA}

\subsection{$\beta\rightarrow\infty$ line}\label{AP:betainf}
To obtain the $\beta\rightarrow\infty$ line we substitute the plane-wave
like solution $\psi_n =\sqrt{d} e^{in\theta}$ with $\theta=\pi$
into Eq. (\ref{eq3}) and obtain
\begin{equation}
\begin{split}
h&=-\frac{\gamma}{\mu^2}\ln(1+\mu d)+ \frac{\gamma}{\mu} d -2d = -\frac{\gamma}{\mu} a+\frac{\gamma}{\mu^2} (e^{\mu a} -1) \\ & -\frac{2}{\mu}(e^{\mu a}-1) ,~~\text{and}\\
a&=\frac{1}{\mu}\ln(1+\mu d).
\label{eq2h}
\end{split}
\end{equation}
\subsection{$\beta=0$ line}\label{AP:beta0}
The grand-canonical partition function for the Hamiltonian Eq.~\eqref{eq3} can be written in the form
\begin{equation}
\textit{Z}= \int_{0}^{\infty} \int_{0}^{2\pi} \prod_{n=1}^N d\phi_n dA_n e^{-\beta (\mathcal{H}+\alpha \mathcal{A})},
\label{eq:Part_funct_Z}
\end{equation}
where parameter $\alpha$ plays the role of chemical potential.
Here $a$ and $h$ can be defined as
$$a=\frac{<\mathcal{A}>}{N}=-\frac{1}{\beta N}\frac{\partial \ln(Z)}{\partial \alpha},$$
and
$$h=-\frac{1}{N}\frac{\partial \ln(Z)}{\partial \beta}-\alpha a.$$\

After integration over the phase variable $\phi_m$ (see Eq.~\ref{eq33}) the following expression is obtained:
\begin{equation}
\begin{split}
Z&=(2\pi)^N\int_{0}^{\infty}\prod_{n=1}^{N} dA_n \\& \times
I_0(2\beta\sqrt{A_nA_{n+1}})\,e^{-\beta \sum_n
	((-\frac{\gamma}{\mu^2}+\frac{\alpha}{\mu}) \ln|1+\mu \, A_n|
	+\frac{\gamma}{\mu} A_n) }.
\label{a4}
\end{split}
\end{equation}

From this expression we find in the limit $\mu=0$ (no non-local
nonlinearity) the whole set of equations derived for the DNLS model
with only local nonlinearity \cite{Rasmussen:2000,Rasmussen:2000:a}.

The line $\beta=0$ which separates the Gibbsian from the non-Gibbsian regime for the Salerno lattice can be obtained in two ways:  

\subsubsection{Method I: Analytical} \label{AP:TIO}
In the limit $\beta \rightarrow 0$, $I_0(2\beta\sqrt{A_mA_{m+1}}) \approx 1$. Now we take $\beta \gamma = x$ and $\beta \alpha = y$, $A=z$. Then Eq.~\eqref{a4} can be expressed as
\begin{equation}
Z=(2\pi)^N E(x,y)^N,
\end{equation}
where
\begin{equation}
E(x,y)=\int_{0}^{\infty} dz e^{x \frac{1}{\mu^2}(\ln|1+\mu z|-\mu z) -y\frac{1}{\mu}\ln|1+\mu z|}.    \label{b2}
\end{equation}
Here we take $y=\beta \alpha \equiv \delta$, a finite quantity. Consequently
\begin{equation}
\begin{split}
a&=-\left. \frac{\partial \ln E(0,y)}{\partial y} \right|_{y=\delta}, \\
h+\alpha \, a&={-} \left. \gamma \frac{\partial \ln E(x,{\delta})}{\partial x}\right|_{x=0}-\left. \alpha \frac{\partial \ln E(0,y)}{\partial y}\right|_{y=\delta}. \label{b6}
\end{split}
\end{equation}
We then obtain
\begin{equation}
\begin{split}
a&=-\left. \frac{1}{E(0,y)}\frac{\partial E(0,y)}{\partial y}\right|_{y=\delta}=\frac{1}{{\delta}-\mu}, \\
h+\alpha a&=\left. {-} \gamma \frac{1}{E(0,{\delta})} \frac{\partial E(x,{\delta})}{\partial x}\right|_{x=0}-\alpha \frac{1}{E(0,y)} \left. \frac{\partial E(0,y)}{\partial y}\right|_{y=\delta} \\
&=\gamma \frac{1}{({\delta}-2 \mu) ({\delta}-\mu)} +\alpha a,\\
h&=\gamma \frac{a}{(\frac{1}{a}-\mu)},
\label{b6}
\end{split}
\end{equation}
We rewrite it as 
\begin{eqnarray}
h=\gamma \frac{a^2}{(1-a \mu)}. \label{anb6} 
\end{eqnarray}

\subsubsection{Method II: Transfer integral operator (TIO) method}\label{AP:TIO}

We apply the TIO method to Eq.~\eqref{a4} which we rewrite as
\begin{equation}
\begin{split}
Z&=(2\pi)^N\int_{0}^{\infty}\prod_{m=1}^{N} dA_m I_0(2\beta\sqrt{A_mA_{m+1}}) \\& \times
e^{-\beta \sum_m (\frac{-\gamma+\alpha\mu}{2\mu^2} (\ln|1+\mu \, A_m|+\ln|1+\mu \,
	A_{m+1}|) +\frac{\gamma}{2\mu} (A_m+A_{m+1})) }. \label{a5}
\end{split}
\end{equation}
In order to evaluate the integral, we consider the thermodynamic limit $N\rightarrow \infty $ of the system and evaluate using the integral equation
\begin{equation}
\begin{split}
\int_{0}^{\infty}\prod_{m=1}^{N} dA_m K(A_m, A_{m+1}) y(A_m) = \lambda y(A_{m+1}), 
\end{split}
\label{eq:kernal}
\end{equation}
where $K(x, y)=I_0(2\beta\sqrt{xy})e^{-\beta (\frac{-\gamma+\alpha\mu}{2\mu^2} (\ln|1+\mu \, x|+\ln|1+\mu \,
	y|) +\frac{\gamma}{2\mu} (x+y)) }$ is the kernel of the integral operator of Eq.~\eqref{eq:kernal}.

Here $K(x,z)$ is symmetric and in the limit $z\rightarrow \infty$,
$\int \int K(x,z)dx\,dz$ should converge.

In the TIO, the partition function can be expressed in a simple
form: $Z\approx (2\pi \lambda_0)^N$, where the largest eigenvalue $(\lambda_0)$
of the kernel function is only taken into account. Therefore,
approximately the norm and energy densities are
\begin{equation}
\begin{split}
a=-\frac{1}{\beta \lambda_0}\frac{\partial \lambda_0}{\partial \alpha},
~~\text{and}~~
h=-\frac{1}{\lambda_0}\frac{\partial \lambda_0}{\partial \beta}-\alpha a.
\end{split}
\end{equation}
\section{The finite time mLCE}\label{lyp_3}
\begin{figure}[!htbp]
	\includegraphics[angle=0, width=\linewidth]{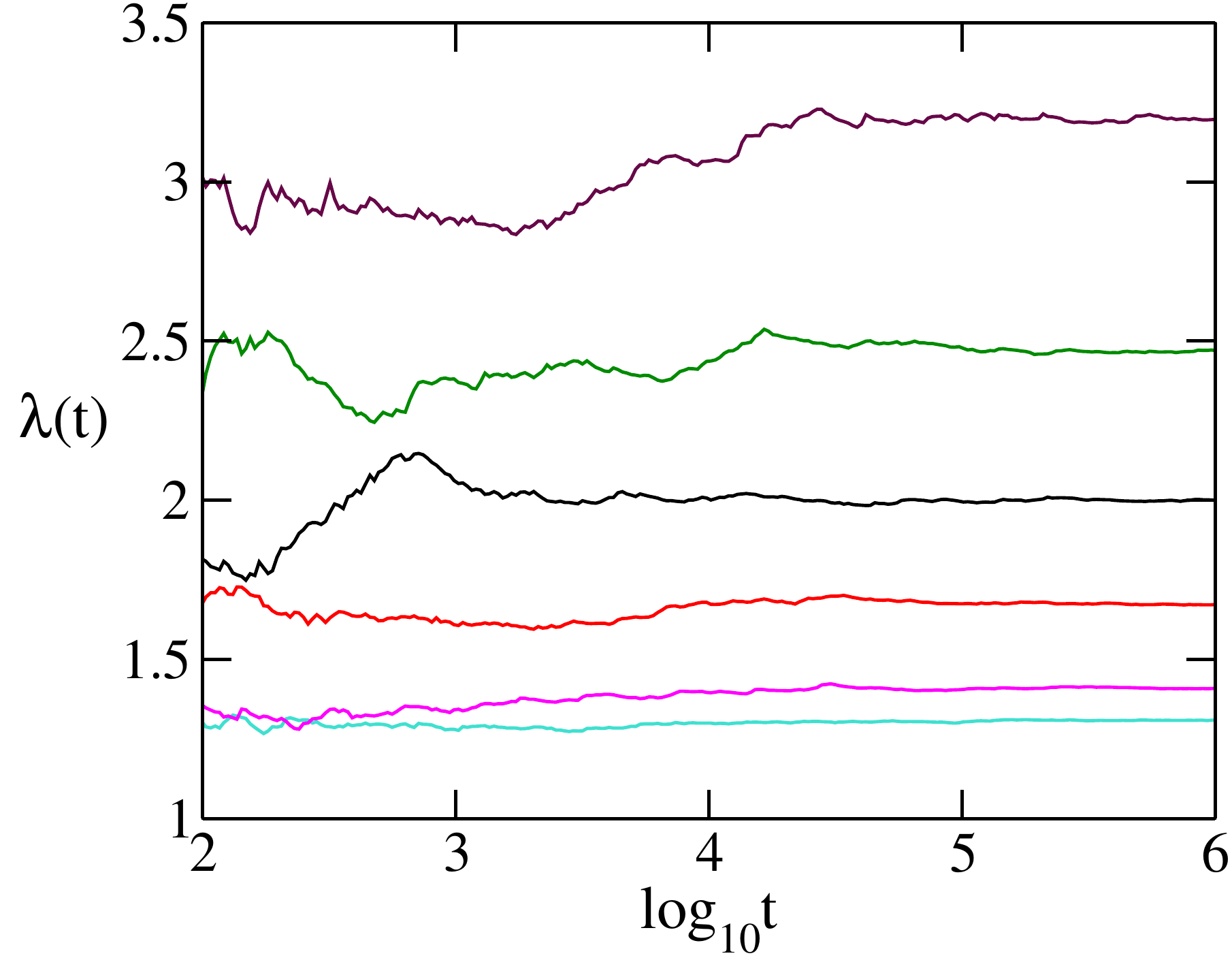}
	\caption{The finite time mLCE $\lambda(t)$ for different $\mu$ values ($\mu = 0.05,~0.1,~0.2,~0.3,~0.4,~0.5$, from bottom to top) and for the parameters $(a=1.5, h=3)$. The value of  $\lambda(t)$ at $t=10^{6}$ is shown in Fig.~\ref{fig:lyp}.} 
	\label{fig:lypt}
\end{figure}
 Fig.~\ref{fig:lypt} shows the finite time mLCE $\lambda(t)$ for different $\mu$ values ($\mu = 0.05,~0.1,~0.2,~0.3,~0.4,~0.5$, from bottom to top) and for the parameters $(a=1.5, h=3)$. The value of  $\lambda(t)$ at $t=10^{6}$ is shown in Fig.~\ref{fig:lyp}. 
%
  \bibliographystyle{apsrev4}
\let\itshape\upshape
\normalem
\bibliography{reference1}
%
\end{document}